\documentstyle[preprint,aps]{revtex}
\tighten
\begin{document}
   
\title{Sufficient Conditions for Apparent Horizons
  in Spherically Symmetric Initial Data}
\author
  {Jemal Guven${}^{(1)}$ \thanks{\tt{ jemal@nuclecu.unam.mx}}
and
   Niall \'O Murchadha${}^{(2)}$\thanks{\tt{ niall@ucc.ie}}}
\address{
${}^{(1)}$\ Instituto de Ciencias Nucleares \\
 Universidad Nacional
Aut\'onoma de M\'exico\\
 Apdo. Postal 70-543, 04510 M\'exico, D.F., MEXICO \\
$^{(2)}$\ Physics Department, University College Cork \\
Cork, IRELAND \\
}
\maketitle

\begin{abstract}
We establish sufficient conditions for the
appearance of both apparent horizons and singularities
in spherically symmetric initial data when spacetime is
foliated extrinsically. Let $M$ and $P$ be respectively the
total material energy and the total material current contained in some
ball of radius $\ell$. Suppose that the dominant energy
condition is satisfied. We show that if $M- P \ge \ell$
then the region must possess a future apparent horizon for some
non -trivial closed subset of such gauges. The same inequality
holds on a larger subset of gauges but with a
larger constant of proportionality which depends weakly on the gauge.
This work extends substantially both our joint work on
moment of time symmetry initial data as well as
the work of Bizon, Malec and \'O Murchadha
on a maximal slice.
\end{abstract}
\date{\today}


\section{Introduction}

This paper is part of an ongoing examination of the constraints in
spherically symmetric general relativity\cite{I,II,III}.
Here we would like to establish sufficient conditions for
the appearance of apparent horizons and singularities in general
initial data. Because of their very different nature
we defer the examination of necessary conditions to
another publication \cite{V}.

Ideally one would like to go about
this in a manifestly covariant way avoiding
the necessity to introduce a gauge. Unfortunately,
this is well beyond our present technical capacity.
We proceed in a canonical way:
the initial data consists of the intrinsic and extrinsic geometry
on some spacelike hypersurface satisfying the
constraints\cite{I}

\begin{equation}
K_R\left[K_R+2K_{\cal L}\right]-
{1\over R^2}\Big[2 \left(R R^\prime \right)^\prime -R^{\prime 2}
-1 \Big]=8\pi \rho\label{eq:ham}\end{equation}
and
\begin{equation}K_R^\prime +
{R^\prime \over R}(K_R-K_{\cal L})=4\pi J\,.\label{eq:mom}\end{equation}
We have parametrized the
line element on the spatial geometry as follows

\begin{equation}ds^2= d\ell^2+R^2 d\Omega^2\,,\label{eq:lineel}\end{equation}
where $\ell$ is the proper radial distance on the surface and  $R$ is
the areal radius. All derivatives are with respect to $\ell$.
In a spherically symmetric spacetime, the extrinsic
curvature is completely characterized by the two scalar functions
$K_{\cal L}$ and $K_R$,
proportional respectively to the velocities of $\ell$ and $R$
normal to the hypersurface. If $n^a$ is the outward pointing
unit normal to the two-sphere of fixed
radius in the hypersurface, we can write

\begin{equation}
K_{ab}= n_a n_bK_{\cal L} + (g_{ab}-n_a n_b)K_R\,.
\label{eq:scalars}\end{equation}
We assume that both $\rho$ and $J$ are appropriately bounded
functions of $\ell$ on some compact support.
We choose to foliate spacetime extrinsically. This involves fixing some
scalar function of the extrinsic curvature tensor, $K_{ab}$.
We focus on the subset of extrinsic time foliations of spacetime of the
form

\begin{equation}
K_{\cal L} + \alpha K_R =0 \,,\label{eq:alpha}
\end{equation}
where $0.5\le \alpha <\infty$
but is otherwise an arbitrary functional of the
initial data, $R$ and $K_R$. $\alpha=0.5$ and $\alpha\to \infty$
define the superspace lightcone.
While this might not be the most general extrinsic time gauge,
a very large degree of freedom is admitted.

If cosmic censorship is valid, the existence of an apparent horizon
provides a natural boundary on the configuration space
between regular data and data that is singular or
will develop a singularity.

We recall that a future (past) apparent horizon
exists when the divergence $\Theta_+(\Theta_-)$ of outward pointing,
future (past) directed null rays vanishes on a closed
surface --- in our case a two -sphere of fixed proper radius.
It is easy to show\cite{BMM2}, for spherical initial data, that if there
exists a
non-spherical trapped surface or apparent horizon then there also must exist
a spherical one.
We can express $\Theta_\pm = \omega_\pm/R$, where

\begin{equation}
\omega_\pm = 2 (R' \pm R K_R)\,,\label{eq:omegapm}
\end{equation}
are the optical scalars introduced in \cite{MOM}.
A future apparent horizon therefore occurs whenever
$\omega_+=0$.
If $\omega_+\le 0$, we say that the surface is future trapped.
The transcription for past horizon will
always be obvious so henceforth will be omitted.
With $\omega_+$ and $\omega_-$, we can reconstruct the
lightcone at each point on the hypersurface (this depends both on the
intrinsic and on the extrinsic geometry). In particular, it is often
useful to cast the constraints in terms of
these variables when we are interested
in identifying apparent horizons \cite{MOM}\cite{I,III}.

When $K_{ab}=0$, the location of an apparent horizon, when one
exists, coincides with an extremal
surface of the spatial geometry \cite{II}. In a
spherically symmetric geometry these are locations where
$R'=0$. In general, apparent horizons do not coincide
with extremal surfaces of the spatial geometry.
Initial data with an apparent horizon
need not even possess an extremal surface and, vice versa.

The spatial geometries we consider consist of a single asymptotically flat
region with topology $R^3$. The appropriate boundary
condition on the metric at the base of the geometry at $\ell=0$ is then

\begin{equation}R(0)=0\,.\label{eq:bc}\end{equation}
We suppose that the center is regular
so that $R^\prime(0)=1$ and $K_R(0)=0$. If no singularity
intervenes between the base and infinity
we will say that the geometry is regular.
In this geometry, the integrated action of the interior
distribution of source energy-momentum can potentially
produce an apparent horizon.

To cast a sufficient condition for the existence
of an apparent horizon we suppose that the
spatial geometry does not possess any future trapped surfaces
and is regular in some bounded region containing the origin
($\omega_+>0$ there).
One needs to then show that
some measure of the material energy content in this region
must be bounded by a measure of the volume of the
region.

The challenge is to identify a useful measure of the material energy
content of a region.
We follow the development of a sufficiency
condition by Bizon, Malec and \'O Murchadha \cite{BMM1,BMM2}
and more recently by Malec and \'O Murchadha\cite{MOM} and \cite{II}.
At a moment of time symmetry,
the natural measure of material energy for
casting a sufficiency condition for an
apparent horizon was
shown to be the material energy, $M$. For general initial data,
the corresponding measure for a future (past) apparent horizon
was shown to be the difference (sum), $M\mp P$, where
$P$ is the total radial material momentum
given by integrating the material
current over the proper spatial volume,

\begin{equation}
(M,P) = 4\pi \int _0^\ell d\ell\, R^2 (\rho,J)
\,.\label{eq:P}\end{equation}
The greater the net outward flux (positive $P$),
the lower $M-P$ --- the more difficult to
form a future apparent horizon.

Clearly, we need to make some assumptions
about matter to proceed. We will assume that
matter satisfies the dominant energy condition in this region:

\begin{equation}
|J| \le \rho\,.
\end{equation}
When the
dominant energy condition  is
satisfied, $M\pm P$ is positive.

Let $\ell$ be the proper radius of this region.
Let the dominant energy condition hold everywhere.
Let $1\le\alpha\le 2$ but be otherwise arbitrary.
Then if

\begin{equation}M \mp P \ge \ell\,,
\label{eq:1}
\end{equation}
the region must contain a future (past) apparent horizon.
This is the central result of this paper.

The inequality (\ref{eq:1}) is particularly impressive
because, when $P=0$,
it coincides  with the moment of time symmetry result
which we know to be sharp.
Unlike the moment of time symmetry scenario where one
could fall back on piecewise-constant density models to guide us,
no such exactly solvable safety net is available here.
Even the analogue of the constant density
star proves to be analytically  intractable when $J\ne 0$.

Bizon, Malec and \'O Murchadha\cite{BMM1,BMM2}
using a maximal slice ($\alpha=2$),
had earlier demonstrated that if the weak energy condition holds and if

\begin{equation}
M- P \ge {7\over 6}\ell
\,,\label{eq:7/6}\end{equation}
the spatial geometry must contain a future trapped surface \cite{BMM1,BMM2}.
The numerical coefficient appearing on the right hand side
is not as good as that appearing in Eq.(\ref{eq:1}).
They showed, however,  that this coefficient
is sharp, by explicitly
constructing a solution with $M-P\ge (7/6 -\epsilon)\ell$ but without
any trapped surface. This solution notably did not satisfy the
dominant energy condition.

The result Eq.(\ref{eq:1}) was first established by
Malec and \'O Murchadha for maximal slices \cite{MOM}.
These earlier results were derived assuming global regularity, a condition
we do not require.

The new idea introduced in \cite{MOM} was
to recast the constraints of the theory
in terms of the optical scalar
variables \cite{MOM}\cite{I,III}.
This permits one to enforce the dominant energy condition rigorously ---
even when $\alpha$ varies
so long as it does not stray too close to the
superspace lightcone.

With maximal slicing alone to go by, one can not be
certain that
the inequality possesses any gauge invariant significance ---
our measures of energy and size, after all,  are not spacetime scalars
so one rightly hesitated before jumping to physical conclusions.
The principal strength of the inequality we have derived
is that it not tied to the maximal gauges;
the result is valid for all slices which remain within
the band $1\le \alpha \le 2$ lying at the center
of the superspace lightcone.
At least within the framework of the extrinsic time foliations,
the result does not appear to be sensitive to a change in foliation.
It is not at all obvious that such a happy
outcome is possible. It is remarkable in view of the global nature of the
problem that we can do this for $\alpha$ which is not constant.

What happens when we move outside this band?
If we relax the gauge to $1\le \alpha <\infty$,
it is still possible to
establish an inequality of this form. However, a constant of proportionality,
$C >1$, is introduced. $C$ will be finite if $\alpha$ is
but diverges linearly as $\alpha\to\infty$. However, this
limit corresponds to the polar gauge which breaks down at a horizon.

If we relax the gauge in the other direction
to cover all apparently valid $\alpha$,
so that $0.5\le \alpha <\infty$, such a  bound no longer exists.
There appears to be a genuine obstruction to casting a strong
statement of sufficiency of the form described above as we
approach the superspace lightcone. We need first to weaken the hypothesis
to obtain an inequality.

One way to do this, taking advantage of dominant energy,
is to assume that the interior
contains neither future nor past trapped surfaces.
--- the cost is that the inequality we obtain
does not distinguish between future and past apparent horizons.
When we do this, we obtain an inequality of the form
(\ref{eq:1}), with some new constant of proportionality,
$C >1$ which depends weakly on the gauge.
In fact, it is finite even when $\alpha=0.5$ everywhere.

To track the potential mischief uncontrolled currents can play, we will
relax the energy condition to the
weak energy condition which places no constraint on $J$.
The weak and the dominant are the two energy conditions
that are relevant in initial data.
The strong energy condition which bounds spatial stresses
involves the dynamical Einstein equations
in addition to the constraints.

With the weak energy condition, we find we do not fare so well.
We again find, with the strong form of the hypothesis, that we can only prove
sufficiency for $1\le \alpha <\infty$. However, the constant $C$
will now depend on $\alpha$ for all $\alpha >1$.
It is mimimized at $\alpha=1$ where it
asumes the optimal value $+1$ ---
the dominant and weak energy results coincide in this gauge.

We also will construct a sufficiency condition for
a singularity in the initial data.
Singular geometries can occur even though both $\rho$ and $J$ are finite.
But the only way that the geometry can become singular is by
pinching off at some finite proper radius from the center.
In Sect.6, we demonstrate that the corresponding inequality for
singularities can be cast as $M \ge
2\ell$ which is  independent of $\alpha$ and $P$.

The paper is organized as follows:
In Sect.2 we focus on the dominant energy condition.
We begin, in Sect.2.1  with a derivation
of bounds on the configuration variables
which hold in regions with a regular center
which do not possess an apparent horizon. These generalize bounds
obtained in \cite{III} for globally regular geometries.
In Sect. 2.2 we derive a useful inequality which we use in
 Sect.2.3 to derive (\ref{eq:1}).
In Sects.2.4 and 2.5, we relax the upper and lower range of $\alpha$
respectively.
In Sect.3, we relax the energy condition.
In Sect.4, we examine some interesting special cases.
In particular, we show that outgoing null fluids cannot form future trapped
surfaces.
In Sect.5 we examine singularities in the
initial data.

\vfill\eject
\section{Apparent Horizons: Dominant Energy}

\vskip2pc
\noindent{\bf II.1 Bounds on the Configuration
Space in Regions without Trapped Surfaces}
\vskip1pc

We consider some finite region containing the origin.
To derive a sufficiency condition in the strong form, we
assume only that this region is regular and
does not possess any future trapped surfaces
($\omega_+ >0$ there) --- we make no assumptions
concerning what happens to the geometry outside, it may contain trapped
surfaces, it may even be singular.

Suppose that the dominant energy condition holds
at every point in this region.
When the dominant energy condition
holds, our experience suggests that the appropriate
variables to exploit are the  optical scalars.

Even though the geometry is regular within this region, we cannot suppose
that it remains so outside. In particular, we cannot impose the
the same boundary conditions we did in \cite{III} for everywhere
regular geometries where we could set $R'\to 1$, and $RK_R\to 0$
at infinity consistent with asymptotic flatness.
Remarkably, it is possible to gain
all the control over $\omega_+$ and $\omega_-$ we will need
without any asymptotic control.

We can combine the two constraints to get  simple equations for the
spatial derivative of $\omega_+$ and $\omega_-$:

\begin{eqnarray}
(\omega_+)^\prime= & -8\pi R(\rho-J) -
{1\over 4R}\Big(\omega_+\omega_- - 4 \Big)+ \omega_+ K_{\cal L}
\,,\label{eq:++}\\
(\omega_-)^\prime=
&-8\pi R(\rho+J) +
{1\over 4R}\Big(\omega_+\omega_- - 4 \Big)-  \omega_- K_{\cal L}
\,.
\label{eq:--}\end{eqnarray}
We can, in turn, combine
Eqs.(\ref{eq:++}) and (\ref{eq:--})
to give

\begin{equation}(\omega_+\omega_-)' = -8\pi R[(\omega_+ + \omega_-)\rho -
(\omega_+ -
\omega_-)J] - (\omega_+ + \omega_-)(\omega_+\omega_- -
4)\,.\label{eq:+-}\end{equation}
It was shown in \cite{MOM} and again in
\cite{I} that if we had initial data
which satisfied the dominant energy condition
and which was regular at both the origin and
infinity then a consequence of Eq.(\ref{eq:+-}) was

\begin{equation}\omega_+\omega_-
\le 4\,.\label{eq:4}\end{equation}
Let us assume instead that we have initial data which is regular at
the origin, satisfies the dominant energy condition
and has no future trapped surface inside
some ball of radius $\ell$. Then inside this radius
Eq.(\ref{eq:4}) holds.  The proof
is very easy. We know that $\omega_+\omega_-  = 4$ at the origin and that it
decreases as
soon as dominant matter is entered. Let us assume that it rises up again to 4.
However, since $\omega_+$ is positive there so also must $\omega_-$ be
positive. Then the dominant energy condition
and Eq.(\ref{eq:+-}) gives us that $(\omega_+\omega_-)' \le 0$ there,
a contradiction. This result is gauge invariant.

An even better result holds if $1\le \alpha<\infty$
in the finite region we consider.
In \cite{III} it was shown that when $\alpha\ge 1$, if
we had globally regular maximal initial data which
satisfied the dominant energy condition then

\begin{equation}-2 \le \omega_+, \omega_-  \le 2\,.\label{eq:---}
\end{equation}
While we need the information at infinity to set the lower bound, we
can get the upper bound just from regularity at the origin. Both $\omega_+$
and $\omega_-$ start out equalling 2 at the origin and then decrease as
one moves away.
Assume that $\omega_+$ increases up to 2 again ahead of
$\omega_-$. However, once it reaches 2, from the dominant energy condition
and

\begin{equation}
(\omega_+)^\prime=  -8\pi R(\rho-J) +
{1\over 4R}\Big(4 + (\alpha-1) \omega_+\omega_- - \alpha\omega_+^2
\Big)
\,,\label{eq:+}
\label{eq:-}\end{equation}
we get that $(\omega_+)' \le 0$, a contradiction.
We obtain the same result for $\omega_-$.
This does not depend on the existence of a horizon.
Note that we only demand $\alpha \ge 1$, we do not require that it be a
constant.

This is not, unfortunately, true if $\alpha <1$ anywhere --- indeed, we
constructed an explicit counterexample in \cite{III}.
If, however, the region possesses neither
future nor past apparent horizons, we can
prove that $\omega_\pm\le \Omega $, where

\begin{equation}
\Omega =
{\rm Max}(2,2/ \alpha_{\rm Min})
\,.\label{eq:alp<1}
\end{equation}
The proof is simply a rerun of that
for $\alpha\ge 1$.

Of course, if we demand  global rather than just local regularity
we can exploit the full range of inequalities
derived in \cite{III}. With weak energy, we have $-1 \le R' \le +1$.
With dominant energy, and $1\le \alpha <\infty$, we have

\begin{equation}
|\omega_\pm|\le 2\,,\label{eq:omega2}
\end{equation}
so that
$R K_R = (\omega_+ - \omega_-)/4$ and, of course,
$R' = (\omega_++\omega_-)/4$ are also bounded. We have

\begin{equation}
|R K_R | \le 1
\label{eq:K1}
\end{equation}
and so also get $-1 \le R' \le +1$.

\vskip2pc
\noindent{\bf II.2 A Useful Inequality when $1\le \alpha  \le \infty $}
\vskip1pc

The optical scalar which
marks the presence of a future apparent horizon is $\omega_+$.
Its spatial gradient is determined by Eq.(\ref{eq:+}).
To obtain the appropriate weighting on the sources,
let us recast Eq.(\ref{eq:+})
as an equation for the spatial gradient of $R\omega_+$:

\begin{equation}
(R\omega_+)^\prime = -8\pi R^2 (\rho- J) +
{1\over 4}\left[
4+
\alpha\omega_+\omega_- + (1-\alpha) \omega_+^2 \right]
\,.\label{eq:Romega}\end{equation}
This equation can be integrated out to any proper radius $\ell$ to give

\begin{equation}
R\omega_+ = - 2(M- P) + 2 \Gamma_+
\,,\label{eq:int}\end{equation}
where
\begin{equation}
\Gamma_+ = \int_0^\ell d\ell\, F_+(\omega_+,\omega_-,\alpha)\,,
\label{eq:Gamma+}\end{equation}
and

\begin{equation}
F_+ =
{1\over 8}
\left[4+ \alpha\omega_+\omega_- + (1-\alpha) \omega_+^2 \right]\,.
\label{eq:F}
\end{equation}
$\Gamma_+$
is the natural optical scalar generalization for
future trapped surfaces of $\Gamma$ introduced
in \cite{II}.
In particular, when $K_{ab}=0$, $\Gamma_+ =\Gamma$.

Let us assume that there are no future trapped surfaces inside some
given radius. For $\alpha\ge 1$,
we then have the crude bound

\begin{equation}
F_+\le {1\over 8} \left[4 + \alpha \omega_+\omega_-\right]\le
{1\over 2} (1 +\alpha)\,.
\end{equation}
We can, however, do much better in this range by
exploiting the control we possess over $\omega_+$ and $\omega_-$,
$0< \omega_+\le 2$ and
$\omega_-\le 2$,  to bound $\Gamma_+$
in a more subtle way.

We are interested in finding an upper bound for
\begin{equation}
F^*_+ = \alpha\omega_+\omega_- + (1 - \alpha)\omega_+^2\,,
\end{equation}
assuming $1 \le \alpha < \infty$, $\omega_- \le 2$ and $0 < \omega_+ \le 2$.

The first thing we observe is that since both $\alpha$ and $\omega_+$ are
positive we maximize $F^*_+$ by setting $\omega_- = 2$. Thus we seek an upper
bound for
\begin{equation}
F^{**}_+ = 2\alpha\omega_+ + (1 - \alpha)\omega_+^2.
\end{equation}
The value of this quartic is zero when $\omega_+ = 0$ and 4 when $\omega_+ =
2$. The maximum of the quartic occurs at
\begin{equation}
\omega_+ = {\alpha \over \alpha - 1}
\end{equation}
and it equals
\begin{equation}
F^{**}_{+\,{\rm max}} = {\alpha^2 \over \alpha - 1}
\end{equation}
there. However, it turns out that $\alpha/(\alpha - 1) > 2$ when $\alpha <
2$ so the maximum of $F^{**}_+$ in that range is 4, independent of $\alpha$.
In other words

\begin{eqnarray}
F^*_+ \le
\cases{ 4 & $1 \le \alpha \le 2$\cr
                {\alpha^2 \over \alpha - 1} & $\alpha > 2$\cr}
\end{eqnarray}

Hence we can derive the useful inequality
\begin{eqnarray}
F_+ \le
\cases{ 1 & $1 \le \alpha \le 2$\cr
       1 + {\alpha \over 8(\alpha - 1)} & $\alpha > 2$\,,\cr}
\label{F_+}
\end{eqnarray}
assuming, of course, $\omega_- \le 2$ and $0 < \omega_+ \le 2$.
This inequality is clearly sharp.

\vskip 2pc
\noindent{\bf II.3 Sufficiency Condition: $1\le \alpha \le 2 $}
\vskip1pc

It is clear, from (\ref{F_+}), that if $\alpha$ lies between 1 and 2 (and,
let us stress again, need not be constant) then we get that $\Gamma_+ \le
\ell$.
Thus it immediately follows from (\ref{eq:int}) that if $M - P >\ell$ then
the region must contain a future trapped surface.

\vskip 2pc
\noindent{\bf II.4 Sufficiency Condition: $1 \le \alpha < \infty$}
\vskip 1pc
It is clear that the inequality no longer holds if $\alpha > 2$.
We have instead the weaker bound for all $\alpha\ge 1$:

\begin{equation}
F\le 1 + H(\alpha)\,,
\label{eq:h1}
\end{equation}
where

\begin{eqnarray}
H(\alpha) =
\cases{ 0 & $1\le \alpha \le 2$\cr
{1\over 8}{\alpha^2\over (\alpha-1)} - {1\over 2} & $\alpha\ge 2$\quad .\cr}
\label{eq:h2}\end{eqnarray}

Let $\langle f \rangle$
represent the average of any
function $f$ over the domain $[0,\ell]$.
We then have

\begin{equation}
\Gamma_+ \le (1 + \langle H(\alpha)\rangle) \ell\,.
\end{equation}
$H$ will be finite if $\alpha$ is bounded.
Asymptotically, however, $H$ grows linearly with $\alpha$.
This reflects the fact that, in this limit,
the gauge is pathological precisely at
a horizon so we would not expect any kind of reasonable estimate.

If $\omega_+ >0$ on any spherical domain which includes the origin then

\begin{equation}
M-P < (1  + \langle H(\alpha)\rangle )\ell \,.
\end{equation}
A constant of proportionality has been introduced which is
given by the spatial average of $1+H$ over the domain  of $\ell$.
If this inequality is violated we must have a future trapped surface in the
domain.

The dependence on $\alpha$ is surprisingly weak.
Clearly, no extra cost is involved in accomodating a variable $\alpha$.
In fact, the inequality does not depend
explicitly on the gradient of $\alpha$.

\vskip2pc
\noindent{\bf II.5 Sufficiency Condition: $0.5\le \alpha < \infty $}
\vskip1pc

If $0.5\le \alpha<1$, the derivation above breaks down.
To bound $\omega_+$, we will asssume that not only is the
interior geometry regular and not contain any future trapped surfaces
but, in addition,  that it contain no past trapped surfaces.
Then Eq.(\ref{eq:alp<1}) holds.
We now  have

\begin{equation}
F \le
{1\over 2} \left(1+ \alpha  + {\rm Max}
\Big({1-\alpha\over \alpha_{\rm Min}}, 0\Big)\right)
\,,\end{equation}
where we exploit the positivity of the quasi-local mass
to bound $\omega_+\omega_-\le 4$ and (\ref{eq:alp<1}) to
bound $\omega_+$.

\vfill\eject

\section{Apparent Horizons: Weak Energy}

\vskip2pc
\noindent{\bf III.1 Bounds on the Configuration
Space in Regions without Trapped Surfaces}
\vskip1pc

Let now us relax the energy condition to weak energy, $\rho\ge 0$.

First, we will again assume only that the geometry in
our region is regular and that $\omega_+ >0$ there.
Without the additional control over
$J$, it is not reasonable to expect $K_R$ to be bounded
in any simple way.
In globally regular geometries,

\begin{equation}
-1 \le R'\le 1\,, \label{eq:R1}
\end{equation}
everywhere\cite{III}. If we do not have global regularity, the
lower bound on $R'$ can be breached.
What we have instead is the weaker bound,

\begin{equation} - {\rm Max}(1, |RK_R|_{\omega_+=0})\le R' \le 1
\,.
\end{equation}
The upper bound is a topological one analogous to that on
$\omega_+$ and $\omega_-$ derived above.
The lower bound is a consequence of the fact that when
$\omega_+>0$, then $R' \ge - RK_R$.
Suppose $R'< -1$ at some point $\ell_0<\ell$.
We know that if $R'< -1$,
$R'' < 0$ so that it will henceforth become more negative \cite{III}.
Thus the mimimum value of $R'$ under these circumstances is the
horizon value as claimed. This is {\it not} a marvellous result ---
but will be sufficient for our purposes.

If we relax the hypothesis to global regularity
we can exploit Eq.(\ref{eq:R1}).

\vskip2pc
\noindent{\bf III.2 Sufficiency Condition: $1\le \alpha < \infty $}
\vskip1pc

Suppose that $\rho\ge 0$,
and $1\le \alpha < \infty$ but is
otherwise unspecified.
We will show that if

\begin{equation}
M -P \ge \langle f(\alpha)\rangle \,\ell
\,,\label{eq:M-Pf}\end{equation}
in some region, where

\begin{equation}
f(\alpha) =1+{1\over2}{(1-\alpha)^2\over 2\alpha-1}
\,,\label{eq:f}\end{equation}
the spatial geometry in this region must possess a future trapped
surface or a singularity.
This exactly reproduces
the condition Eq.(\ref{eq:7/6}) when $\alpha=2$.
However, the minimum of
$f(\alpha)$ is assumed when $\alpha=1$ where we reproduce Eq.(\ref{eq:1}).
Curiously, the gauge providing the best bound when we do not assume
dominant energy is not maximal slicing. The likely reason for this
is that in this gauge, $K_R= P/R^2$. Here,
weak energy does as well as dominant energy.

The original proof in \cite{BMM1,BMM2}
for $K=0$ exploited conformal coordinates. Our
approach eschews tying ourselves to any particular
spatial coordinate. Not only
is the  end result independent of the spatial coordinate, it is clear
that the  coordinate invariant approach is not only more transparent
but also more efficient. We will work with the
metric variables $R$ and $K_R$ --- when dominant energy is
relaxed, the advantage we gain from exploiting
the optical scalars is lost.
When Eq.(\ref{eq:alpha}) holds we can rewrite Eq.(\ref{eq:ham})
in the form

\begin{equation}
4\pi  \rho R^2  + \big(RR^\prime\big)'=
{1\over 2}\Big(1+(R^\prime)^2\Big)+{1-2\alpha \over 2}R^2 K_R^2
\,.\label{eq:ham1}\end{equation}
We integrate from $\ell=0$ up to the boundary at proper radius $\ell$:
\begin{equation}
M +RR^\prime=  \Gamma +
\int_0^{\ell} d\ell\, \left({1-2\alpha \over 2}\right) R^2 K_R^2
\,,\label{eq:proof1}\end{equation}
where $\Gamma$ is defined by

\begin{equation}
\Gamma ={1\over 2}\int_0^{\ell} d\ell\,
\Big[1+(R^\prime)^2\Big]
\,.\label{eq:Gamma}\end{equation}
We now eliminate $R^\prime$ in the surface term
in favor of the optical scalar $\omega_+$
and $K_R$ using the defining relation (\ref{eq:omegapm}).

To eliminate the $K_R$ dependence on the boundary which comes along
with the replacement of $R'$ by $\omega_+$,
we note that we can integrate the momentum constraint,
Eq.(\ref{eq:mom}) to obtain \cite{1}

\begin{equation}
R^2  K_R=  P
+\int_0^\ell d\ell\, (1-\alpha)\,R R^\prime  K_R
\,.\label{eq:identity1}\end{equation}
We do not need an explicit solution of the
momentum constraint. In fact, the only explicit use
we will make of the momentum constraint is
Eq.(\ref{eq:identity1}) which does
not depend on the possible spatial variation of $\alpha$.

Substituting Eq.(\ref{eq:omegapm}) and (\ref{eq:identity1}) into
(\ref{eq:proof1}) we now obtain

\begin{equation}
M- P +2\omega_+ =
\Gamma +
{1\over 2}\int _0^\ell d\ell\, (1-2\alpha)\,R^2 K_R^2
-\int _0^\ell d\ell\, (1-\alpha)\,R R^\prime  K_R
\,.\label{eq:proof2}\end{equation}
Let us label the second and third terms on the right hand side,
$I_1$ and $I_2$ respectively.
When $\alpha>0.5$, $I_1$ is manifestly negative.
As such we could discard it to cast (\ref{eq:proof2}) as an
inequality. However, it is clear that we can do better
by first completing the square in the sum of $I_1$ and $I_2$
before discarding:

\begin{equation}
I_1 + I_2 = {1\over 2}
\int _0^\ell d\ell\, (1-2\alpha)\,\Big(R K_R
-{1-\alpha\over 1-2\alpha} R^\prime \Big)^2
+{1\over2}
\int_0^{\ell}d\ell\, {(1-\alpha)^2\over 2\alpha-1}\, R^{\prime2}
\label{eq:proof3}\end{equation}
Now, if  $R^{\prime 2}\le 1$, we have

\begin{equation}
I_1 + I_2
\le  {1\over2}\left\langle{(1-\alpha)^2\over 2\alpha-1}\right\rangle\,
\ell \,.
\label{eq:proof33}\end{equation}
In addition, under these conditions, we obtain the upper bound on $\Gamma$,

\begin{equation}
\Gamma \le \ell
\,.\label{eq:Gamma1}\end{equation}
This is, however, only the case when we assume that the geometry is
regular everywhere.
Let us suppose that $R'=-1$ at some value $\ell_0 < \ell$.
Then, we can decompose the right hand side
of Eq.(\ref{eq:proof2}) into a part
coming from the integration over the domain
$[0,\ell_0]$ and a remainder from $[\ell_0,\ell]$:
The integrand, $\Phi$, appearing in this
latter contribution can be  bounded as follows

\begin{equation}
\Phi \le |RK_R|{}_{\omega_+=0}
(1 + |\alpha -1| - \alpha )
\,,\label{eq:proof4}\end{equation}
which is bounded by zero if $\alpha\ge 1$, so that we can discard it.
We conclude that if the region contains no future
trapped surfaces and its geometry is regular then

\begin{equation}
M -P < \left(1 +
\left\langle{1\over2}{(1-\alpha)^2\over 2\alpha-1}\right\rangle\right)\,\ell
\,.\label{eq:M-P}\end{equation}
This completes the proof.

If we extend the range of $\alpha$ to values below $+1$,
not surprisingly, we need to
relax the hypothesis. However we decide to do this, we appear to
require control over the form of the exterior geometry.
One might try to weaken the hypothesis to permit us
to assume  that the entire spatial geometry
(from its base out to infinity) is free of future trapped surfaces.
However, there is no tactical advantage to this ---
unlike the moment of time symmetry analysis where this
does imply that the geometry is non-singular, here a
past apparent horizon could turn up in the exterior with a singularity
lurking beyond it.

We need to assume that the
entire geometry possesses neither future
nor past horizons or weaker, is regular
not just within the region of interest.
The result will be that one cannot conclude that the apparent
horizon or singularity lies necessarily within the region
when the inequality holds --- it could exist beyond it.

Under these conditions, we again obtain Eq.(\ref{eq:M-Pf}) with the
same function $f(\alpha)$.
We note that $f$ diverges as we
approach the minisuperspace lightcone, $\alpha = 0.5$ and
$\alpha\to \infty$.
This occurs because the discarded negative
term blows up at these two values.
At the former  value, a more ingeniously constructed proof
is likely to remove this infinity. At the latter value,
however, the divergence is a genuine signal of the breakdown of
the gauge.

\vskip 2pc
\noindent{\bf III.3 Sufficiency Condition: $0.5 \le \alpha < \infty$}
\vskip 1pc
We have encountered genuine obstructions
to finding strong sufficient conditions for the
existence of  future horizons as one approachee the superspace lightcone.
Interestingly, one can find a very simple  sufficient condition which is valid
for the whole range of $\alpha$'s if we do not distinguish between future and
past horizons.

Let us consider a region around the origin, and let us assume that there are
no trapped surfaces in this region. We immediately get that in this region we
have
\begin{equation}
0 < R^\prime \le 1\,,
\end{equation}
just assuming that the weak energy condition, $\rho \ge 0$, holds.

Let us consider Eq.(\ref{eq:ham1})
and integrate it out to any $\ell$ in the region with no trapped surfaces.
We replace $(R^\prime)^2$ by 1, and using the facts that
$RR^\prime > 0$ and that
$(1-2\alpha) R^2 K_R^2 \le 0$ we get that
\begin{equation}
M < \ell\,.\end{equation}
Therefore we have shown that if the weak energy condition holds and if $M \ge
\ell$ then we must have a trapped surface within this sphere.
This result is an old moment-of-time-symmetry result in disguise. We do not
need
to assume absence of trapped surfaces to get the inequality, all we need to
assume is the absence of a minimal surface, i.e., $R^\prime \ge 0$, to get that
$M < \ell$.  Therefore we have shown that if $M \ge \ell$ then there must be a
minimal surface within this sphere. Of course, a minimal surface must be a
future
or a past trapped surface.

We can even do better! The term
$(1-2\alpha) R^2 K_R^2/2$ which we threw away can be brought over to
the other side and reunited with the $4\pi  \rho R^2$ term. This combination
is nothing but one quarter of the three scalar curvature, which must be
positive
by the combination of the requirement that $\alpha \ge 0.5$ and the weak energy
condition. Therefore we have recovered the result that if we have a spherical
manifold with positive scalar curvature and if the volume integral of the
scalar curvature within a sphere is greater than $16\pi\ell$, where $\ell$ is
the proper radius of the sphere, then there must be a minimal surface within
this sphere \cite{II}.

\section{Special Cases}

Let us now examine some extreme cases: for
simplicity we will only consider
dominant energy.

Let us consider the two extreme distributions
satisfying the average dominant energy condition
everywhere, but only just, so that $P\to \pm M$ \cite{2}.
If  $|J|= \rho$,  these are
respectively the cases of a radially outward and a radially inward
moving null fluid. In the case $P\to M$,
Eq.(\ref{eq:1}) becomes a vacuous
statement, not surprisingly, as it is
impossible
to form an apparent horizon under such
circumstances.
All we need to do is look at Eq.(\ref{eq:+}) and see that if $\rho = J$ and
if $\omega_+$ is small then $(\omega_+)' \simeq 1/R > 0$. Thus $\omega_+$
can never pass through zero from above and since it starts off equalling 2, it
must remain positive.
In the latter
case, we have that if $2M \ge \ell + \langle H(\alpha)\rangle \,\ell$,
the geometry will possess a future trapped surface. It is twice as
easy to form an apparent horizon with an inflowing null fluid as it is with a
stationary fluid.

It is also possible to tighten the sufficiency condition in the same
way we did in \cite{II}
for $K_{ab}=0$ when $\rho^\prime\le 0$ if, in addition,
we demand that $J$ has a fixed sign. What we do
substitute for $R'^2$ in Eq.(\ref{eq:Gamma1})
using the definition of the quasilocal mass. We have

\begin{eqnarray}
m &=& 4\pi\int_0^\ell d\ell\, R^2 \left[\rho R' + J RK_R\right]\nonumber\\
  &=& {4\pi\over 3}\rho R^3 +
       4\pi\int_0^\ell d\ell\, R^2\left[ J RK_R -
 \rho' R \right]\nonumber\\
  &\ge& {4\pi\over 3}\rho R^3 +
4\pi \int_0^\ell d\ell\, R^3 J K_R\\
&\ge& {4\pi\over 3}\rho R^3\,.\label{eq:msurf}
\end{eqnarray}
The inequality on the second last line obtains when $\rho'\le 0$.
If
$J$ is positive (or negative) everywhere
(In the gauges we are considering $K_R$ possesses the same sign as $J$)
we are left with a sum of two positive terms and so the last inequality
holds.
We now have

\begin{equation}
R'^2 = 1 - {2m/R} + R^2 K_R^2\,\,\le\,\,
      1 - {4\pi\over 3}\rho R^2 + R^2 K_R^2\,,\label{eq:inter}
\end{equation}
so that

\begin{equation}
\Gamma \le \ell- {M\over 3} + {1 \over 2}\int_0^\ell d\ell\, R^2 K_R^2
\,.\label{eq:Gamma2}\end{equation}
Let us examine how this modifies Eq.(\ref{eq:M-Pf}).
The last
term on the right can now be added to the (negative) term of the same form in
Eq.(\ref{eq:proof2})
before the completion of the square. We get

\begin{equation}
{4M\over 3}- P <
{3 + \left\langle\alpha\right\rangle \over 4}\ell
< \left\langle\alpha\right\rangle \ell \label{eq:4M/3-P}\end{equation}
in the range $\alpha\ge 1$.
As before, this is minimized when $\alpha=1$ and when $P=0$
again reproduces the result at a moment of time symmetry. We note,
however, that while the
left hand side  has been improved, the right hand side is weaker.

>From one point of view, Eq.(\ref{eq:4M/3-P}) is not
very satisfactory --- we have broken the symmetry between
$J$ and $\rho$. However, this very asymmetry apparently permits us
to write down a non-vacuous sufficiency condition when
dominant energy is saturated with  $P=M$. Whereas
Eq.(\ref{eq:M-P}) is vacuous under these conditions,
Eq.(\ref{eq:4M/3-P}) provides
the non-trivial statement: suppose $\rho^\prime\le 0$ and the motion
of matter is outward and null, then if

\begin{equation}
M \ge 3 \langle \alpha \rangle \,\ell
\,,\label{eq:Jout}\end{equation}
the spatial geometry will possess an apparent horizon.
This apparently contradicts the proof of the absence of apparent horizons
when one has a
null fluid given in the beginning of this section. The resolution of this
paradox, as we will demonstrate in the
next section, is that Eq.(\ref{eq:Jout}) can never be satisfied.

In a forthcoming publication we will examine
intrinsic time foliations. What is very encouraging is
that essentially the same condition appears
despite the very different nature of the foliation\cite{VI}.

\section{Singularities }

Whereas the sufficiency conditions for apparent horizon were
clearly strengthened by imposing the dominant energy condition,
there is no obvious useful way to import dominant energy into
the statement of a sufficiency condition for singularities.

We have an obvious generalization of the moment of time symmetry
result. We recall that, in general,

\begin{equation}
M +RR^\prime=  \Gamma +
\int_0^{\ell} d\ell\, \left({1-2\alpha \over 2}\right) \,R^2 K_R^2
\,.\label{eq:sing1}\end{equation}
Let us suppose that the geometry is regular everywhere.
Let $\rho$ be positive and $\alpha\ge 0.5$.
Now $\Gamma$ is bounded by one.
Furthermore, $R^\prime\le 1$ so that $R(\ell)\le \ell$ everywhere
on a non-singular geometry and $R^\prime \ge -1$. The  surface term
is therefore bounded from below by $-\ell$.
Finally, the second term on the RHS is negative  ---
The $K_{ab}$ dependence is trivially handled. Thus  we get

\begin{equation}
M \le 2\ell \,.\label{eq:sing2}\end{equation}
The inequality is independent of the value of $\alpha$
so long as it is bounded from below by $0.5$ --- we
do as well as we did at
a moment of time symmetry.
As at a moment of time symmetry, if we place constraints on the
sources
it is possible to tighten the inequality. We note that when
$\rho^\prime\le0$
and $J$ is positive (or negative) everywhere, we can
exploit Eq.(\ref{eq:Gamma2})  to get

\begin{equation}
M \le {3\over 2}\ell
\,.\label{eq:sing3}\end{equation}
Thus we see the vacuous nature of Eq.(\ref{eq:Jout}).

Unlike the analogous configurations which
occur at a moment of time symmetry. which are non - singular,
$\rho'\le 0$ is not sufficient to guarantee a non-singular
geometry --- the geometry can
still turn singular if $J$ is large enough.
We therefore cannot claim that
Eq.(\ref{eq:sing3}) represents a universal bound when $\rho^\prime <0$
and $J$ is positive (negative).

The singularity condition is not  symmetrical in $M$ and $P$.
Unlike the apparent horizon conditions, where $P$ shows up
in the combination $M\pm P$, it does
not arise in a natural way in
either Eq.(\ref{eq:sing2}) or Eq.(\ref{eq:sing3}).

\section*{Acknowledgements}
We gratefully acknowledge support from
CONACyT Grant 211085-5-0118PE to JG and Forbairt Grant SC/96/750 to
N\'OM

\vfill\eject


\begin{references}
\bibitem{1}
The privileged role of the gauge with $\alpha=1$ is evident.
\bibitem{2}
We could alternatively consider matter satisfying the
weak energy condition locally but satisfying the
averaged strong energy condition globally.
Local violations of the dominant energy condition are possible.
\end{references}
\end{document}